\documentclass[%
 aip,
 rsi,%
 amsmath,amssymb,
reprint,%
]{revtex4-1}

\usepackage[final]{graphicx}
\usepackage{dcolumn}
\usepackage{bm}

\begin{document}

\preprint{AIP/123-QED}

\title[Measurement of THz Faraday angle]{Simultaneous measurement of circular dichroism and Faraday rotation\\ at terahertz frequencies using heterodyne detection}

\author{G. S. Jenkins}
\author{D. C. Schmadel}%
\author{H. D. Drew}
\affiliation{%
Center for Nanophysics and Advanced Materials\\Department of
Physics, University of Maryland, College Park, Maryland 20742, USA
}%
\date{\today}

\begin{abstract}
A far-infrared system measures the full complex Faraday angle,
rotation as well as ellipticity, with an unprecedented accuracy of
10\,$\mu$rad/T. The system operates on several far-infrared laser
lines in the spectral range from 0.3 to 6 THz and produces results
as a continuous function of temperature from 10 to 310K and
applied fields between $\pm$ 8\,T. Materials successfully measured
include GaAs 2-DEG heterostructures, various high temperature
superconductors including Bi$_2$Sr$_2$CaCu$_2$O$_{8+x}$, Pr$_{2 -
x}$Ce$_{x}$CuO$_4$, and La$_{2-x}$Sr$_x$CuO$_4$, and single
crystals of the topological insulator Bi$_2$Se$_3$.

\end{abstract}

\pacs{78.20.Ls,78.66.Bz,74.25.Gz,74.72.-h,}

\keywords{Faraday effect,  ellipticity, circular dichroism, Terahertz, heterodyne detection}
\maketitle

\section{\label{sec:Introduction}Introduction}

Hall effect measurements are a powerful tool routinely used in
condensed matter physics. Extension of these measurements to
optical frequencies provides dynamical characterization of the
off-axis conductivity $\sigma_{xy}$ where optical propagation is
in the z direction. This off-axis conductivity when combined with
the  longitudinal conductivity $\sigma_{xx}$ from standard optical
measurements produces the full complex conductivity tensor along
with the Hall angle defined as
$\tan(\theta_H)=\sigma_{xy}/\sigma_{xx}$.

This additional information available from infrared (IR) Hall
measurements has been particularly useful in investigating
strongly interacting electron systems, materials often defying
paradigms in condensed matter physics established in the twentieth
century. Since interaction effects can enter into $\sigma_{xy}$
differently than
$\sigma_{xx}$,\cite{SchmadelPRBRapid,Kontani_2008Review} the IR
Hall effect provides new insights to strong correlation physics.
Similar to $\sigma_{xx}$, Hall spectroscopy probes the energy
scales of the system including energy gaps and carrier scattering
rates. It also obeys sum rules that permit a more global view of
the effects of the interactions on the electronic structure of
materials.\cite{drew_sumrule_1997,SchmadelPRBRapid,zimmers-IRHall-2007}
IR Hall measurements have led to several recent important
findings: refuting the spin-charge separation scenario for the
anomalous behavior of the dc Hall effect in
cuprates,\cite{SchmadelPRBRapid,Cerne1,Cerne2} firmly establishing
that the anomalous dc Hall effect in the paramagnetic state of
n-type and p-type cuprates is caused by current vertex corrections
produced by electron-electron scattering mediated by
antiferromagnetic fluctuations, \cite{Jenkins2009PRB,
Jenkins-overdopedPCCO, JenkinsBi2212} observing the first evidence
of small Fermi pockets in underdoped cuprates,\cite{rigalPRL2004}
demonstrating that inelastic scattering enters into the Hall
conductivity differently than the longitudinal conductivity,
\cite{Cerne1,Cerne2} and characterizing the Fermi pockets formed
by a spin density wave gap in underdoped
Pr$_{2-x}$Ce$_{x}$CuO$_4$.\cite{Jenkins2009PRB,
zimmers-IRHall-2007}

Hall effect measurements at very low optical excitation energies
in the THz spectral region probe intrinsic properties associated
with the Fermi surface (FS) which are sensitive to the FS topology
as well as Fermi velocity and scattering rate anisotropies.
\cite{Jenkins2009PRB, Jenkins-overdopedPCCO, JenkinsBi2212} One of
the advantages of the far-IR (FIR) Hall measurements over the
mid-IR (MIR) Hall measurements\cite{RSICerne} is that for many
material systems, there are known intermediate energy scales
between 10 and 100 meV which make comparison between MIR Hall
measurements difficult with dc transport and other low frequency
probes.

Important future applications include extending quantum Hall
effect (QHE) measurements to the THz regime. The recent appearance
of novel 2-D systems such as graphene and the conducting surface
states on topological insulators (TI) has stimulated interest in
the transport and magneto-transport in systems with ``massless"
Dirac electronic dispersion. Topological insulators are predicted
to form a new 2-D quantum state of matter fundamentally related to
the edge states in the QHE.\cite{QiZhangPRB2008}  Although many
exotic properties are predicted, of particular interest are
signatures at finite frequencies of a quantum Hall step
$\sigma_{xy}= (1/2) e^2/h$ -- even though there exist no edge
states -- never before observed. For that matter, the frequency
dependence of the QHE has never been measured in any material at
optical frequencies, a topic we are currently pursuing. In
particular, the Faraday angle in both graphene and TIs is
predicted to be quantized in units of the fine structure
constant.\cite{QiZhangPRB2008,maciejko_topological_2010,morimoto_Aoki_2010}

The instrument to be described in the following pages is capable
of measuring  the full complex Faraday angle defined as:
\begin{equation} \label{eqn; defn Faraday Rotation}
\tan  \theta_F  \equiv  \frac{t_{yx}}{t_{xx}}
\end{equation}
where $t_{yx}$ and $t_{xx}$ are the transmission amplitudes
through the sample, which is situated in a magnetic field
perpendicular to its surface and parallel to the propagation
direction of the optical beam. The system operates on various
laser lines in the terahertz spectral region (0.3 to 6 THz) and
produces data to an unprecedented accuracy of 10\,$\mu$rad/T and
as a continuous function of temperature from 10 to 310\,K. Using
the full complex Faraday angle, together with independent
measurements of the longitudinal conductivity $\sigma_{xx}$, one
may then calculate both the the complex Hall angle and
$\sigma_{xy}$.

In what follows we shall first present an overview of the system
followed by a detailed examination of several of the more critical
components and design issues. Finally, we discuss the system
calibration and operation. Useful mathematical notation and models
appear in the appendices.

\section{\label{sec:InstrumentDescription}Instrument overview}

\begin{figure}
  \centering
  \includegraphics[scale=1]{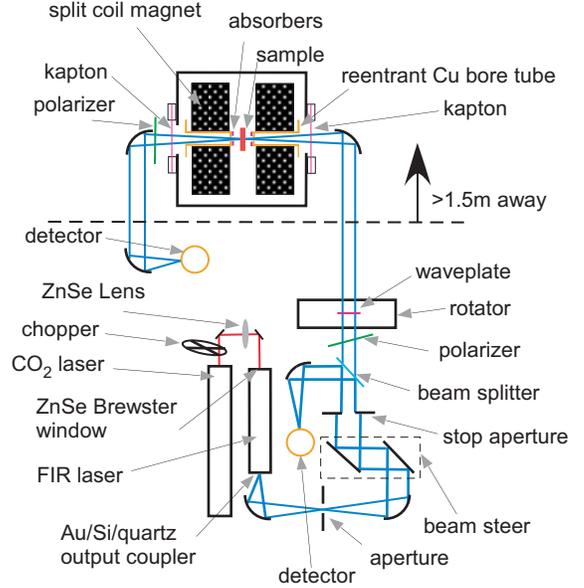}
  \caption[Diagram of Faraday experiment]{(Color online) Schematic of the complex Faraday angle measurement system operating on various laser lines in the THz region.}\label{fig Optical Table and ELectronics Layout}
\end{figure}

Fig.\,\ref{fig Optical Table and ELectronics Layout} presents a
schematic representation of the system. At the beginning of the
optical path, a chopped, continuous wave CO$_2$ laser pumps a
single longitudinal mode, far-infrared (FIR), molecular vapor
laser which is the source of the THz radiation. An aluminum
parabolic mirror focuses the output of the FIR laser onto a
pinhole aperture to remove spatial noise. A second aluminum
parabolic mirror collimates the emerging beam and directs it into
a beam steering assembly consisting of two flat mirrors. Next, an
aperture stop reduces the beam diameter to a size so that the
radiation does not encounter the sides of the bore tube of the
mechanical rotator. A thin mylar sheet beam splitter splits off a
small portion of this reduced diameter beam for use as a power
reference. The main portion of the reduced diameter beam passes
through the mylar beam splitter and enters the bore of the high
speed mechanical rotator. A waveplate mounted within the bore
typically spinning at $\sim$3000\,RPM modulates the beam so as to
introduce various harmonics of the rotation frequency into the
polarization states. A third aluminum parabolic mirror focuses the
beam onto the sample located in a $\pm$8 Tesla superconducting
Helmholtz magnet. The magnetic field is aligned parallel to the
beam path.

In the Faraday geometry, the portion of the beam transmitted by
the sample passes through an analyzer polarizer and onto a fourth
aluminum parabolic mirror, which re-collimates it. A fifth
aluminum parabolic mirror focuses the beam into the output
detector. Note that the beams leading to and from the magnet and
the fourth and fifth mirrors are collimated. This allows
increasing the separation between the magnet and the detector and
laser source so as to reduce the effects of the strong magnetic
field upon these components. Also note that, except for the small
amount of spatial noise introduced by the aperture stop prior to
the mechanical rotator, the system operates essentially within a
single spatial mode. This limits noise from modal interference
whose frequencies resulting from vibration, thermal drifting, etc.
would otherwise fall within the mechanical rotator frequency and
its harmonics.

An ideal rotating waveplate transmits linearly polarized light
when either the ordinary or extra-ordinary axis aligns with the
polarizer. This occurs four times per revolution of the waveplate
giving rise to a fourth harmonic signal. With a moments thought,
one will recognize that a sample inserted in the magnet so as to
cause a real Faraday rotation Re$(\theta_F)$ will change the
relative phase of this fourth harmonic. In fact, the fourth
harmonic phase change is directly proportional to the Faraday
rotation. An angular position sensor on the mechanical rotator
allows determination of this phase change.

Considering no sample in place, the polarization state oscillates
between pure left and right handed circular polarized light twice
upon one full rotation of an ideal waveplate. Upon introduction of
a sample, one can see qualitatively that the second harmonic
signal is then sensitive to the induced ellipticity or circular
dichroism. The appendix presents, in a very manageable notation, a
complete derivation of all the various signals one might expect
for typical samples. Corrections to accommodate errors and losses
in the various system components is accomplished via an \textit{in
situ} calibration technique discussed in Sec.\,\ref{sec; Sample
polarizer calibration}.

In general, a sample induces polarization rotation, characterized
by Re$(\theta_F)$, as well as ellipticity, characterized by
Im$(\theta_F)$. The resulting signals recorded at the output
detector include dc level $R_{dc}$ of the transmittance through
the sample measured at the chop frequency $\omega_{chop}$ of the
CO$_2$ pump laser and $A_{i,2\omega}$, $A_{o,2\omega}$,
$A_{i,4\omega}$, and $A_{o,4\omega}$, which are the second and
fourth harmonic signals on the output detector that are in-phase
(i) and out-of-phase (o) with the original harmonic signal that
results when no sample is present or when the magnetic field is
zero. By taking appropriate ratios of $A_{i,2\omega}$,
$A_{o,2\omega}$, $A_{i,4\omega}$, $A_{o,4\omega}$, and $R_{dc}$,
one may extract the complex Faraday angle via
Eq.\,\ref{eq:detectorsignaHS}. Note that similar to heterodyne
detection schemes, for small angles the desired quantities are
proportional to the phase and therefore the transverse $t_{xy}$
amplitude rather than being proportional to the power which is not
so accurately measured for small values.

\section{Critical components and design issues}
\subsection{Etalons and Multiple Reflections}

Multiple reflections within optical elements and between optical
elements can result in spurious signals and must be minimized.
Consider, for example, that the Faraday effect accumulates
regardless of the direction of the light beam. Whereas a single
pass through a sample or optical element may produce a rotation of
$\theta$, a pass forward, then backward, then again forward will
produce a rotation of $3\theta$. This is a concern not only for
the sample but for any optic within the magnetic field, for which
reason we use thin polyimide film, ~25 micron, as the magnet
vacuum windows. Even outside of the magnetic field multiple
reflections can cause small variations over time due to very small
motions or vibrations of the hardware like flexing of the magnet
cryostat or flexing of the the aluminum optical table from cryogen
boil-off, motion of the floor, thermal drift, etc.

To control these effects, optical elements may be: chosen to be
much thinner than the wavelength; conditioned with an
antireflection (AR) coating; and/or set at an angle to the optical
path. For example, a dielectric AR layer was utilized on an
LaSrGaO$_4$ (LSGO) substrate to impedance match between vacuum and
the FIR substrate index of $n \approx 4.4$. The matching layer was
a piece of z-cut quartz chemically etched to a thickness equal to
an odd multiple of a quarter wavelength and held against the
substrate by means of a spring. The refractive index of quartz in
the FIR is 2.12 which is close to the geometric mean, 2.10, of
vacuum and LSGO. High resolution FTIR transmission measurements,
corresponding to various laser frequencies\cite{Spizler} were used
to confirm the etched thickness.

Another type of broadband AR coatings that may be applied to
sample substrates, waveplates, and some detector cold filters is a
thin metallic film.\cite{DrewARcoatings} The Fresnel reflection
coefficient for an interface between vacuum and a thin metallic
film on a dielectric of index n is $r=(n-1- Z_0 /
R_\square)/(n+1+Z_0 / R_\square)$ which is zero under the
condition that the sheet resistance $R_\square=Z_0/(n-1)$ where
$Z_0=377 \,\Omega$ is the impedance of free space. $R_\square$ is
approximately constant provided the scattering rate is
substantially greater than the optical frequency, a condition met
in the FIR for nichrome (NiCr) films. Such AR coatings on quartz,
silicon, gallium arsenide (GaAs), and LSGO were characterized via
high resolution FTIR transmission measurements and used
successfully in a number of Faraday experiments.

In some very specific circumstances, etalons within the sample can
be used to emphasize particular material properties. For example,
appropriate engineering of the etalon in TIs allows isolation of
the surface state response producing a Faraday signal which
exactly equals the fine structure
constant.\cite{maciejko_topological_2010} However, in most
circumstances, elimination of the etalon is desirable to simplify
the optical system and subsequent analysis.

\subsection{Sample design and mounting considerations}

Ideally the sample should be made as large as possible and the
laser spot as small as possible. However, diffraction limits the
minimum spot size at the sample to $1.22 \lambda f/D$. For our
magnet cryostat, $f/D$ is 3.5 so the spot size for 20\,$cm^{-1}$
radiation is 2\,mm. Samples much smaller than the spot size can be
located within an aperture to reduce the background. However, an
aperture introduced at this point in the beam path will cause
spurious polarization changes at its edges that can mask the
Faraday signals. Constructing the aperture from an absorbing
material like graphite reduces such effects.

\begin{figure}
  \centering
  \includegraphics[angle=0, scale=.25, keepaspectratio=true]{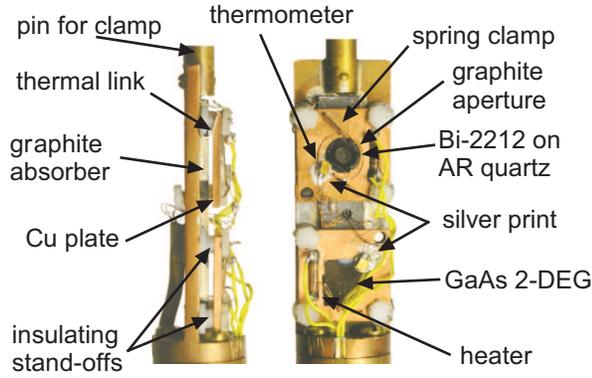}
  \caption[Picture of end of sample stick]{(Color online) Photographs of the sample stick cold finger depict two views. Total length shown is $\sim$2.5 inches. The top sample is peeled single crystal Bi$_2$Sr$_2$CaCu$_2$O$_{8+x}$ (Bi-2212) on an AR coated quartz substrate. The bottom sample is a NiCr coated GaAs 2-DEG heterostructure with an AR coating.}
  \label{fig;samplestickBothViews}
\end{figure}

Many of the samples studied were highly reflective and poorly
transmissive. By strategically placing graphite absorbers on the
re-entrant copper bore tubes (Fig.\,\ref{fig Optical Table and
ELectronics Layout}) and sample stick
(Fig.\,\ref{fig;samplestickBothViews}), and mounting the sample on
an aperture, the effects of beam scatter were measured to be $\sim
5\times 10^{-5}$ which is negligible in nearly all circumstances.
Imaging the sample onto a spatial filter placed at the location of
the detector has been used effectively to further attenuate
scattering effects. Note that an aperture in this location will
not introduce any spurious signal by modifying the polarization
since the analyzer polarizer has already decoded the polarization
information.

For the simple case of thin film samples, the Faraday angle is
related to the Hall angle as shown in Eq.\,\ref{eq; Faraday To
Hall Angle Conversion} by a factor involving the sheet resistance
$1/(\sigma_{xx} d)$ acquired by an independent FTIR spectroscopic
measurement. This value may be engineered to be close to 1 by
choosing the thickness of the film. However, as shown in
Eq.\,\ref{eq; Faraday To Hall Angle Conversion}, the sheet
resistance should be as small as possible while still allowing
adequate transmission. As a gauge, samples which attenuated the
throughput intensity by four orders of magnitude were successfully
measured (like the Bi$_2$Sr$_2$CaCu$_2$O$_{8+x}$ sample deep in
the superconducting state as shown in
Fig.\,\ref{fig:Bi2212}).\cite{JenkinsBi2212} The signal at this
level is detector noise limited.

The sample mounting system must maintain the sample position as
well as control the sample temperature. Its central component is a
sample stick, which was modified from a design originally made by
Oxford Instruments. The cold finger shown in
Fig.\,\ref{fig;samplestickBothViews} is cooled by a continuous
flow of liquid helium. The sample rests on a 1/16" thick copper
plate which has a graphite aperture smaller than the sample. A 100
$\Omega$, 1/4 W metal film resistor epoxied into a slot in the
sample plate serves as a heater. To restrict the motion of the
sample during temperature and magnetic field sweeps, a 3" long
thin-walled stainless-steel tube is silver soldered to the end of
the cold finger. This tube fits into a teflon insert in an
aluminum chuck which we added to the cryostat housing. The chuck
is actuated from outside the housing. A copper wire serves as a
thermal link whose size is adjusted to allow sufficient cooling of
the sample while throttling heat flow to the sample stick from the
sample heater minimizing thermal expansion effects. The
temperature can be continuously varied from 10 to 310 K. Thermal
stresses on the sample are minimized while maintaining adequate
thermal coupling to the copper plate by mounting the sample with a
spring clip on one end and silver epoxy on the other. A Lakeshore
340 temperature controller monitors and regulates the temperature
of the sample and simultaneously monitors the temperature of the
sample-stick.

\subsection{Magnet}

Our magnet is an Oxford Instruments Helmholtz or split coil
superconducting magnet that can scan magnetic field between
$\pm$8\ T in approximately 12 minutes. The magnet is maintained in
a liquid helium bath. The vacuum windows are 0.001" thick
polyimide which are slightly permeable to helium, allowing a small
amount of exchange gas to enter the vacuum chamber. This causes
cooling of the polyimide windows resulting in water condensation.
This is mitigated by blowing room temperature air constantly
across the windows. The entire magnet dewar rests on four 1/4-20
screws to allow height and tilt adjustment for aligning to the
optical beam. An Oxford model PS120-10 power supply controls the
superconducting magnet current.

\subsection{Laser system}

As mentioned earlier, the laser system is a molecular vapor FIR
laser pumped by a continuous wave CO$_2$ laser. The CO$_2$ laser
is a model PL5 from Edinburgh Instruments Ltd. and is tunable from
9 to 11\,$\mu m$. Its output is a 7.5\,mm diameter beam with
maximum power of 50\,W. The output is focused by an AR-coated ZnSe
lens into the cavity of the FIR laser through a small hole in one
of the FIR cavity mirrors.

The FIR laser is a home-built dielectric waveguide
laser\cite{FIRLaserdesignpaper1, FIRLaserdesignpaper4} with two
adjustable flat mirrors on either end. The input coupler is a 2
inch diameter molybdenum flat mirror with a 1.5\,mm aperture
purchased from SPAWR Industries. The input coupler accommodates a
ZnSe Brewster window which minimizes destabilizing back-reflection
feedback into the CO$_2$ laser. The output coupler is a quartz
substrate with a silicon dielectric coating grown to a thickness
matching a 1/4 wave at 10\,$\mu$m to provide a high reflectivity
for the pump beam. A gold annulus sputtered on the silicon surface
defines a 1\,cm diameter FIR output aperture. The total area of
the output coupler allows the CO$_2$ beam to be maximally
maintained within the FIR laser cavity while concurrently
filtering out MIR frequencies from optical elements downstream.
The large aperture minimizes diffraction effects. A 1.3\,m fused
silica tube forms the FIR waveguide and maintains its length. Its
low thermal expansion coefficient $7.5 \times 10^{-7} / K$ helps
to maintain stability. Diffraction losses strongly attenuate
higher order modes \cite{FIRLaserdesignpaper9} while lower order
modes are only slightly attenuated due to the grazing incidence
with the
waveguide.\cite{FIRLaserdesignpaper6,FIRLaserdesignpaper1,
FIRLaserdesignpaper2, FIRLaserdesignpaper3, FIRLaserdesignpaper4}
The power output depends strongly on the chosen FIR frequency, but
is typically between 1 and 50mW. The axial mode spacing is $\Delta
\nu_{ax} = c / (2 L) \sim 150$ MHz, the  Doppler broadening is
$\sim 5$ MHz for methanol, and pressure broadening is typically
$\sim 40 \text{MHz/torr} \approx 4$ MHz under the operating
pressure of 100\,mTorr. Typical gasses used are CD$_3$OD
(24.6\,$cm^{-1}$), CH$_3$OH (84.7\,$cm^{-1}$), CH$_2$F$_2$
(42.3\,$cm^{-1}$), and CH$_3$OD (175.4\,$cm^{-1}$). A continuous
flow at typically $\sim$100\,mTorr pressures is maintained, and a
gas handling system with a cold trap is used to recapture the
gasses.

The FIR laser lines are scanned for the presence of multiple
resonances using a step scan fourier transform polarizing
spectrometer. The cyclotron frequency of a calibrated GaAs 2-DEG
provides a convenient \textit{in situ} tool to ascertain the FIR
laser wavelength (see Sec.\,\ref{sec; waveplate calibration}).

\subsection{Mechanical Rotator}
The high speed mechanical rotator spins an optic of up to 1"
diameter at an adjustable speed up to 3000\,rpm resulting in a
rotator fundamental frequency of 50\,Hz. The optic is located
within a nominal 1" diameter threaded bore tube. Two threaded
rings sandwich the optic between two o-rings that provide
cushioning to distribute stress which would otherwise introduce
non-uniformities into the optic. The bore tube is pressed into two
ABEC-7 angular contact ball bearings which are themselves pressed
into the aluminum rotator housing. An Animatics model SM2315D
servomotor with programmable PID feedback drives the bore tube by
means of a kevlar reinforced polyurethane timing belt which
engages a pulley pressed onto the bore tube end. An intermediate
belt tensioner reduces belt slap and jitter. Black paint on a
portion of the bore tube pulley provides a target for an optical
sensor which supplies the electronic phase reference for the
lockin amplifiers. The phase noise as measured on the lockins is
within $1:10^4$ (at 50\,Hz using a 200\,msec time constant).

An array of various thicknesses of x-cut quartz first-order
waveplates are chosen to cover the entire spectral region from 3
to 240\,$cm^{-1}$. Whenever the thickness of the waveplate is not
optimal resulting in a retardance of $\Delta \beta\neq\pi/2$,
signal is sacrificed (see Eq.\,\ref{eq:detectorsignaHS}). The
limiting upper frequency when using quartz waveplates is $\sim 240
\, cm^{-1}$.

\subsection{Detectors, mirrors, polarizers, and beam splitter}
The detectors are silicon composite bolometers purchased from
Infrared Laboratories. The primary detector is cooled to 2\,K and
the reference detector to 4\,K. The 77K and helium cold stage
filters are quartz, the vacuum window is wedged white polyimide,
and a Winston cone collects radiation up to an f/\# of 3.5. The
characteristic frequency response of the primary detector is
$\approx$500\,Hz with a noise-equivalent-power of $2 \times
10^{-13}$ W/Hz$^{1/2}$ at 100\,Hz and a responsivity of $2\times
10^5$\,V/W.

All 90 degree off-axis parabolic mirrors are 4" aluminum
substrates with MgF$_2$ coatings purchased from Janos technology.
Flat mirrors are bare aluminum evaporated on optically flat glass.

Polarizers are evaporated Au on 6\,$\mu$m thick mylar with
4\,$\mu$m pitch and 2\,$\mu$m  linewidth purchased from
ScienceTech. These are mounted taut between aluminum rings. Their
measured extinction coefficient is $10^{-5}$ at 84.7\,$cm^{-1}$.

The reference beam splitter is a 0.00048" thick mylar sheet that
is mounted taut between 2 flanges with o-rings. The reflectance is
$\sim$10\% at 84.7\,$cm^{-1}$ and is slightly frequency dependent.

\subsection{Electronics}

The bolometer detectors are equipped with pre-amplifiers that
allow a high impedance ($\gg$ 50 $\Omega$) output at gain settings
of 1, 200, or 1000.

The output signal of the transmission bolometer detector is fed
into an EG\&G 5113 low-noise pre-amplifier which is then fed into
a 7280 EG\&G lock-in amplifier. A homemade optical encoder reads
the frequency of the rotator bore and generates TTL pulses to the
lock-in amplifier. The 7280 decomposes the signal into in- and
out- of phase components of the second and fourth harmonics. An
EG\&G 7260 and a EG\&G 7265 lock-in amplifier measure the
magnitude of the chopped signal from the reference and
transmission detectors.

The detectors are nonlinear causing sums and differences of
frequencies of various noises and signals to appear in the
spectrum. A spectrum analyzer enables choosing $\omega_{chop}$ and
$\omega$ (typically 311 and 49\,Hz, respectively) such that the
frequencies of interest do not inadvertently overlap with other
harmonics or their sums and differences.

An in-house Labview program simultaneously controls and acquires
data from the magnet power supply, temperature controller, EG\&G
5113 preamplifier, and three lock-in amplifiers.

\begin{figure}\centering
 \includegraphics[scale=.90, keepaspectratio=true]{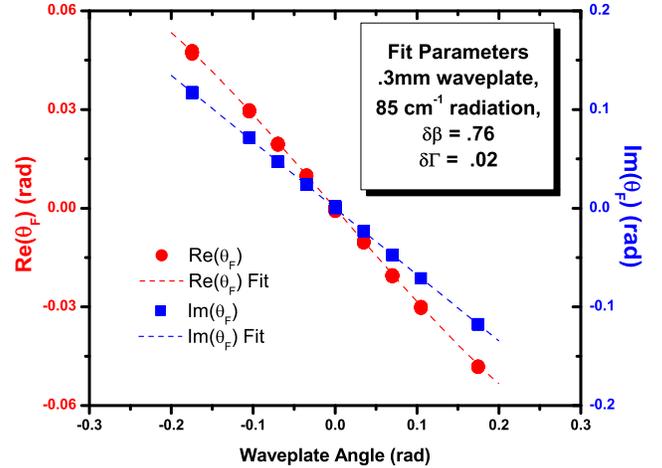}
 \caption[Im($\theta_F$) calibration with quartz waveplates]{(Color
 online) The real and imaginary part of the Faraday angle is measured as a %
  function of angle for a quartz waveplate and fit using the waveplate parameters %
   $\delta \beta$ and $\delta \Gamma$. The imaginary part is calculated using the %
   fit parameters and overlayed on top of the data.}\label{fig; 84 and 175 WP Calibration}
\end{figure}

\section{\label{sec:InstCalibration}Instrument calibration and Operation}

\subsection{Electronics}
\subsubsection{Lock-in amplifiers: harmonic analysis and setting the
phase}

The complete detector signal is given by
Eq.\,\ref{eq:shorthanddetectorsignaHS} and
\ref{eq:detectorsignaHS}. The EG\&G lock-in amplifiers measure the
RMS-amplitude of all signals. Optically chopping produces a square
wave whose first harmonic amplitude is 0.637 compared to the
square wave amplitude. The measured $R_{dc}$ value needs to be
multiplied by a factor of 2 since the actual dc level is the
peak-to-peak value of the square waveform. In this case, the
magnitude of the chopped signal measured by the lock-in amplifier
needs to be multiplied by $2 / .637$ to be used in
Eq.\,\ref{eq:detectorsignaHS} as $R_{dc}$.

The phase of the $4\omega$ lock-in is set to zero in zero magnetic
field. However, inspection of Eq.\,\ref{eq:detectorsignaHS} shows
that the $2\omega$ phase can not accurately be set under the same
conditions since both the in- and out- of phase signals are close
to zero (since $\Delta \Gamma \approx 0$). Conceptually, the
$2\omega$ phase can only be set when there exists a substantive
Im($\theta_F$) signal such that $A_{o,2\omega}>>A_{i,2\omega}$.
Since $\Delta \Gamma$ is small, maximizing Im($\theta_F$) by
adjusting the phase of the $2\omega$ signal in magnetic field is,
in principle, a good method for setting the phase. In practice, a
single lock-in is used to measure both the $2\omega$ and $4\omega$
signals. The phase of the second harmonic is set equal to the
phase of the fourth harmonic. In software, the phase of the
$2\omega$ channel is then rotated by $\Delta \phi_{2\omega}$ to
maximize the Im($\theta_F$) signal. $\Delta \phi_{2\omega}$ is
found to be the same for a variety of samples and corresponds to
the phase shifts observed when characterizing the frequency
dependence (where phase information was acquired) of the detector
between the $2\omega$ and $4 \omega$ frequencies.

\subsubsection{Detector frequency roll-off}
The relative phase and amplitude response of the detectors and
preamplifiers as a function of modulation frequency are
experimentally characterized. One method involves double chopping
an FIR beam whereby one chopper continuously sweeps frequency
while the other remains at fixed frequency. The amplitude response
is used to correct the detector signal between the three relevant
modulation frequencies: $2\omega$, $4\omega$, and $\omega_{chop}$.

\subsection{$\Delta \beta$ and $\Delta \Gamma$ from FTIR
spectroscopy}\label{sec;FTIR-WP-calibration} FTIR spectroscopy was
performed on waveplates inserted at various angles $\theta$
between two parallel polarizers. The transmitted intensity
(relative to the input intensity) is given by the following:
\begin{equation*}
\begin{split}
I(\theta) = & \,e^{-2 \gamma_1}\cos^4 (\theta) +  e^{-2
\gamma_2}\sin^4 (\theta) \\
& + 2 e^{-(\gamma_1 + \gamma_2)}  \cos (\Delta
\beta)\sin^2(\theta)\cos^2(\theta)
\end{split}
\end{equation*}
where $\gamma_1$ and $\gamma_2$ are the absorption along the e-
and o- axis of the waveplate, and $\Delta \beta$ is the
retardance. Taking ratios of I($\theta$) for various angles
$\theta$, $\Delta \Gamma=\gamma_2-\gamma_1$ and $\Delta \beta$ as
a function of frequency can be derived.

\subsection{$\Delta \beta$ and $\Delta \Gamma$ \textit{in situ}
calibration: polarizer as sample}\label{sec; Sample polarizer
calibration} The preferred method to calibrate waveplates is
\textit{in situ} in which the unknown waveplate is inserted into
the rotator. Adjusting a polarizer to various angles $\alpha$
inserted immediately downstream from the rotator yields a detector
signal given by Eq.\,\ref{eq:polarizerAsSampleDetSigs}. After
taking ratios of the various amplitudes, there are a total of four
equations. If we treat as unknowns the four quantities $\Delta
\beta$, $\Delta \Gamma$, and the frequency roll-off of the
detector between $2 \omega$ and $4 \omega$ and between $2 \omega$
and $\omega_{chop}$, then simultaneously solving for all four
quantities gives an \textit{in situ} calibration of the waveplate
parameters and frequency roll-off of the detector. Small
polarization errors introduced from imperfections associated with
the waveplate (for example, an imperfect AR coating) and/or
induced effects from the kapton vacuum windows are corrected with
this method by giving an effective $\Delta \beta$ and $\Delta
\Gamma$ which is always within $\sim 5\%$ of independent
characterizations of the waveplate. The detector frequency
roll-off never requires correction of more than small fractions of
a percent. An analysis justifying this method of error correcting
is detailed in Ref.\,\onlinecite{GJThesis}.

The real part of the Faraday angle measured by rotating the
polarizer to small angles ($\pm 5$\,degrees) by Eq.\,\ref{eq;
RotWP-HS-smallangle} exactly equals $\alpha$ (given by
Eq.\,\ref{eq:polarizerAsSampleDetSigs}) thus providing a very good
absolute calibration of Re($\theta_F$).

\begin{figure}
\centering
\includegraphics[scale=.90,keepaspectratio=true]{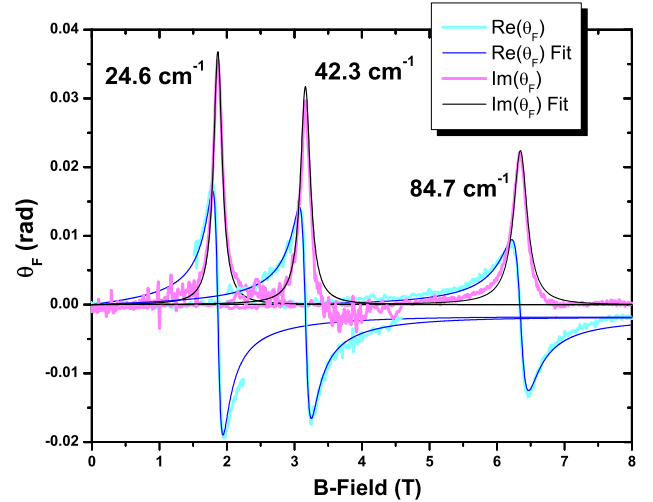}
\caption[Re($\theta_F$) and Im($\theta_F$) calibration with GaAs
2-DEG ]{(Color online) The Faraday angle measured at 3 frequencies
as a function of magnetic field for an AR-coated GaAs 2-DEG
heterostructure with a 25 $\Omega / \Box$ layer of NiCr at 80 K.
The real part of the Faraday angle is fit using the mobility
($\mu$), electron density (n), and NiCr electron scattering rate
($\gamma_{NiCr}$) as parameters. For all frequencies, $n=1.78
\times 10^{15} m^{-2}$ and $\gamma_{NiCr} = 5100 \; cm^{-1}$. The
mobility was found to be 14, 12, and 8.4 $m^2 / V sec$ for 24.6,
42.6, and 84.7 $cm^{-1}$, respectively. The GaAs electron mass was
taken to be $0.07 \times m_e$.} \label{fig; GaAs 2-DEG
Calibration}
\end{figure}

\subsection{Calibration of Im($\theta_F$): inserting a waveplate
or 2-DEG as a reference sample}\label{sec; waveplate calibration}

By inserting a waveplate  immediately downstream from the rotator
at various angles $\alpha$ while measuring Re($\theta_F$) and
Im($\theta_F$) given by Eq.\,\ref{eq:ReImFA-sampleWP}, it is
possible to calibrate the imaginary part of the Faraday angle with
respect to the real part. The inserted waveplate parameters
$\Delta \beta'$ and $\Delta \Gamma'$ are obtained by fitting the
real part of the Faraday angle. The fit parameters are then used
to calculate the imaginary part, and overlayed on the imaginary
Faraday angle data. Results for the 84.7 $cm^{-1}$ data using a
0.3 mm thick waveplate is shown in Fig.\,\ref{fig; 84 and 175 WP
Calibration}. Waveplates measured at other frequencies were
checked in the same manner. The fit parameters of the waveplate
agree with other independent waveplate characterizations
(Sec.\,\ref{sec;FTIR-WP-calibration} and \ref{sec; Sample
polarizer calibration}).

A GaAs 2-DEG sample is used for calibration of the Im($\theta_F$)
as well. Im($\theta_F$) and Re($\theta_F$) have a well defined
relationship as shown in Eq.\,\ref{eqn; Drude complex FA} allowing
calibration of Im($\theta_F$) with respect to Re($\theta_F$). A
GaAs 2-DEG heterostructure with an AR coating and a 25 $\Omega /
\Box$ layer of NiCr on the opposite side of the wafer was
deposited in order to attenuate the transmission level to $\sim
1\%$. The transmission level and magnitude of the Faraday signals
are similar to many of our experiments. The mobility and the
density of electrons are used as fit parameters. The small Faraday
response of the NiCr layer is accounted for with a simple Drude
model where the electronic scattering rate is a fit parameter.

This sample was used to check the imaginary part of the Faraday
angle with respect to the real part at three separate frequencies:
24.6, 42.3, and 84.7 $cm^{-1}$. The real part was fit, and the
imaginary part was then calculated using the fit parameters from
the real part. The results are shown in Fig.\,\ref{fig; GaAs 2-DEG
Calibration}.

The same electron density and NiCr scattering rate were used for
all three frequencies, namely $1.78 \times 10^{15}\, m^{-2}$ and
$5100 \, cm^{-1}$. The mobility is a function of frequency and
found to be 14, 12, and 8.4 m$^2$/ V sec at 24.6, 42.3, and
84.7\,$cm^{-1}$, respectively. The temperature was held at 80\,K
for all data. Since the mobility is inversely proportional to the
effective mass as well as the scattering rate, the data suggest
that either the mass or the hall scattering rate is enhanced as
the frequency is increased. Since the resonant frequency is not
significantly shifted away from the calculated cyclotron frequency
($\propto 1/m$), the frequency dependence is associated with the
scattering rate.

All of the data show some small asymmetry about the resonance in
the imaginary part of the Faraday angle. This might be a result of
some small amount of mixing between the real and imaginary parts
of the Faraday angle. Imperfect sample AR coatings will cause
etalons leading to some small amount of mixing.

\section{Instrument performance: measurements on Bi-2212}
To illustrate the performance of the instrument, we show reported
results\cite{JenkinsBi2212} on optimally doped single crystal
Bi$_2$Sr$_2$CaCu$_2$O$_{8+x}$ (Bi-2212) samples 100\,nm thick with
an area defined by an aperture 2.5\,mm in diameter as shown in
Fig.\,\ref{fig;samplestickBothViews}. The thin cuprate
superconducting crystal was mounted to an AR coated quartz
substrate. The longitudinal conductivity $\sigma_{xx}$ was
independently measured via FTIR spectroscopy.

\begin{table}\centering
\begin{tabular}{|l|l|l|l|l|l|}\hline\hline
\multicolumn{2}{|c}{\rule [-3mm]{0mm}{8mm} \bfseries 24.6
$cm^{-1}$} & \multicolumn{2}{|c}{\bfseries 42.3 $cm^{-1}$}&
\multicolumn{2}{|c|}{\bfseries 84.7 $cm^{-1}$}
\\ \hline %
\multicolumn{1}{|c}{\rule [-3mm]{0mm}{8mm} \bfseries
Re($\theta_F$)} & \multicolumn{1}{c}{\bfseries Im($\theta_F$)} &
\multicolumn{1}{|c}{\bfseries Re($\theta_F$)} &
\multicolumn{1}{c}{\bfseries Im($\theta_F$)} &
\multicolumn{1}{|c}{\bfseries Re($\theta_F$)} &
\multicolumn{1}{c|}{\bfseries Im($\theta_F$)} \\ \hline \hline %
2.031 & 0.374    & 1.992 & 0.457   &  1.865  & 1.081
 \\
1.937 & 0.409   &  2.074 & 0.450    & 1.886  & 1.105
\\
2.057 & 0.394 & 1.992 & 0.457  &  1.957 &  1.067 \\
1.982 & 0.373 & 2.035 & 0.446 & 1.911   &
1.097 \\
-.--- & -.--- & 2.051 & 0.443
 & -.--- & -.---\\ \hline
 2.002   & 0.388  & 2.029  & 0.451  & 1.905 & 1.088 \\ \hline \hline
\end{tabular}
\caption{Slopes of individual magnetic field sweeps in mrad/T
performed on Bi-2212 at 100\,K like those depicted in
Fig.\,\ref{fig:Bi2212}(a,b). The last row is the column
average.}\label{tbl; slopes of B-field sweeps}
\end{table}

\begin{figure*}[!t]
\includegraphics[scale=1,clip=true]{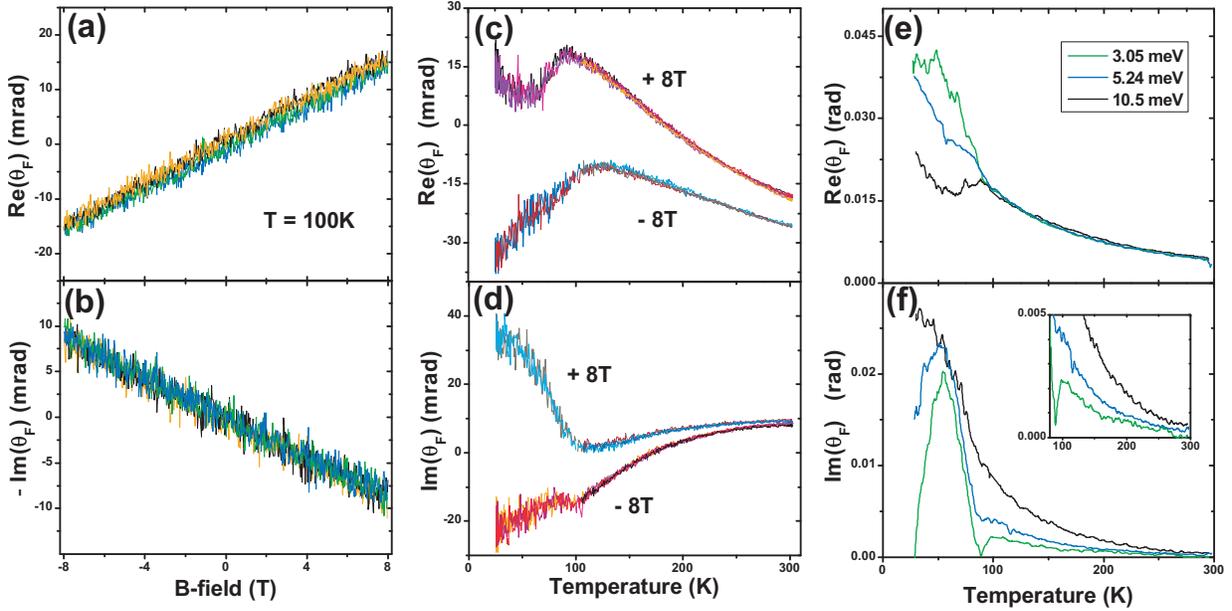}
\caption{(Color online) (a,b) Faraday angle measured at 10.5\,meV
(84.7\,cm$^{-1}$) and 100\,K versus B-field. Four scans are
depicted. (c,d) Faraday angle measured at $\pm8$\,T versus
temperature. Ten scans are depicted. (e,f) Faraday angle versus
temperature found by subtracting scans in $\pm8$\,T, averaging
multiple scans together, and performing a 3\,K rolling average
with temperature.} \label{fig:Bi2212}
\end{figure*}

Magnetic field sweeps to $\pm 8$\,T were performed at 100\,K.
Fig.\,\ref{fig:Bi2212}(a,b) shows the Faraday angle at 100\,K as a
function of magnetic field for 10.5 meV (84.7 $cm^{-1}$)
radiation. A compilation of the slopes associated with individual
magnetic field scans of both the real and imaginary part of the
Faraday angle for all frequencies is presented in Table \ref{tbl;
slopes of B-field sweeps}.

Temperature sweeps from 100 to 300\,K and from 300 to 25\,K are
shown in Fig.\,\ref{fig:Bi2212}(c,d). The transmission changes by
two orders of magnitude between high and low temperature since the
sample becomes superconducting. Evident in the traces is the
growing level of detector noise as temperature is decreased. There
exists a temperature dependent background produced by sample
movement. Subtracting scans in $\pm$8\,T and dividing by 2 nulls
the background signal. The resulting Faraday angle is shown in
Fig.\,\ref{fig:Bi2212}(e,f) where multiple scans were averaged and
then smoothed with a 3\,K moving average. Magnetic field sweeps
taken at various discrete temperatures give the same value of the
Faraday angle as the temperature sweep data.

The current limitation of the signal-to-noise level is imposed by
drift produced by sweeping temperature and to a lesser extent from
sweeping magnetic field. The current noise from these effects is
0.1\,mrad, or 10$\mu$rad/T.

Variations of the system are currently underway. dc current
measurements which depend upon the chirality of the incident light
are planned which utilize the same optical polarization modulation
technique.\cite{moore_orenstein_2009} A Kerr geometry is being
configured since the technique is sensitive to a single surface,
particularly important in measuring topological insulators.

\begin{acknowledgments}
The authors extend their thanks to Andrei B. Sushkov, Geoff Evans,
and Jeffrey R. Simpson for their assistance in performing the
various reported measurements, and Genda Gu and Matthew Grayson
for supplying the Bi-2212 and GaAs 2-DEG samples. This work was
supported by the CNAM, NSF (DMR-0030112), and DOE
(DE-AC02-98CH10886).
\end{acknowledgments}

\appendix
\section{\label{sec:Appendix}Mathematical formalism}
\subsection{General Notation}
The entire optical system can be described as a series of transfer
functions which operate on the initial state of an incident
electro-magnetic plane wave.

Dirac notation is used to efficiently keep track of the
polarization state as is described in
Ref.\,\onlinecite{SchmadelThesis} and Ref.\,\onlinecite{RSICerne}.
Under our convention whereby the radiation propagation is in the
positive z direction, the electric polarization vector $|E
\rangle$ can be represented in either a linear (L) or circular (C)
polarization basis as:
\begin{tabular}{ccc}
~~~~~~~~~~~$ \langle L|E \rangle$ = $\begin{pmatrix}
    E_x \\
    E_y
\end{pmatrix}$
& and & $ \langle C|E \rangle$ = $\begin{pmatrix}
    E_+ \\
    E_-
\end{pmatrix}$\\
\end{tabular}\\
where $E_{\pm}=E_x \hat{x} \pm i E_y \hat{y}$. $E_+$ then
represents the circular polarization with the electric field
rotating in a positive direction about the z axis in the lab
frame. The main types of optical components of our system are:
polarizer (P), waveplate (WP), and sample (S), which may be
represented as follows:
\begin{tabular}{lcr}
~~~~~~~~~~~~~$\langle L|P|L \rangle$ &=&  $\begin{pmatrix}
    1   &   0   \\
   0    &   0
\end{pmatrix}$\\[2.0ex]

~~~~~~~~~~~~~$\langle L|WP|L \rangle$ &=& $\begin{pmatrix}
    e^{ \mathit{i} \beta_1 - \Gamma_1 }   &   0   \\
   0    &   e^{ \mathit{i} \beta_2 - \Gamma_2 }
\end{pmatrix}$\\[2.0ex]

~~~~~~~~~~~~~$\langle L|S|L \rangle$ &=& $\begin{pmatrix}
   t_{xx}   &   t_{xy}   \\
   t_{yx}    &   t_{yy}
\end{pmatrix}$
\end{tabular}\\
$\Delta \beta = \beta_1 - \beta_2$ and $\Delta \Gamma=\Gamma_1 -
\Gamma_2$ are the differential phase shift and absorption between
linearly polarized light along the ordinary and extra-ordinary
optical axis of the waveplate. For an ideal quarter waveplate
$\Delta \beta = \frac{ \pi }{2}$ and $\Delta \Gamma = 0$. Actual
values are experimentally determined by an \textit{in situ}
calibration technique (Sec.\,\ref{sec; Sample polarizer
calibration}) and verified by FTIR measurements
(Sec.\,\ref{sec;FTIR-WP-calibration}). $t_{xx}$, $t_{xy}$,
$t_{yx}$, and $t_{yy}$ are the transmission amplitudes for an
arbitrary sample. Most of the measured samples are rotationally
invariant meaning that $t_{xx} = t_{yy}$ and $t_{xy} = -t_{yx}$.
In this case:
\begin{equation*} \label{eqn; defn rotationally invariant sample}
\begin{tabular}{lcr}
$\langle L|S_{0}|L \rangle$ &= & $
\begin{pmatrix}
   t_{xx}   &   t_{xy}   \\
   -t_{xy}    &   t_{xx}
\end{pmatrix}$\\[1.0ex]
\end{tabular}
\end{equation*}

The rotation of a component by an angle $\phi$ may be represented
by an arbitrary angle $\phi$ clockwise by the following operator:

$ \langle L| R( \phi ) |L \rangle =
\begin{pmatrix}
\cos  \phi     &    -\sin  \phi    \\
\sin  \phi   &    \cos  \phi  \\
\end{pmatrix}$\\

A rotating quarter-waveplate spinning at an angular frequency
$\omega$ about the positive z axis is represented by:
\begin{equation*}\label{equ; rotating waveplate matrix}
\langle L|RWP( \omega )|L \rangle =
 \langle L|R(  \omega t)|L \rangle  \langle L|WP|L \rangle \langle L| R( -\omega t ) |L \rangle
\end{equation*}

\subsection{Detector Signal Analysis}
\subsubsection{Rotationally invariant Sample}

Using the above notation, the intensity at the output detector of
the system shown in Fig.\,\ref{fig Optical Table and ELectronics
Layout}, for a rotationally invariant sample is:
\begin{align}\label{eq:shorthanddetectorsignaHS}
I_D &= | \langle L|P|L \rangle \langle L|S_{0}|L \rangle \langle
L|RWP( \omega  )|L \rangle \langle L|P|L \rangle \langle L|E_0
\rangle  |^2 \notag\\
&=R_{dc}+A_{o,2\omega} \sin  2 \omega t + A_{i,2\omega} \cos 2
\omega t\\
&~~~~+A_{o,4\omega} \sin  4 \omega t + A_{i,4\omega} \cos  4
\omega t \notag
\end{align}
where,
\begin{align}\label{eq:detectorsignaHS}
&R_{dc}=e^{ -\Gamma} |t_{xx}|^2 \frac{1}{4} ( (1 - |\tan
\theta_F|^2 ) \cos \Delta
\beta + \notag \\
&~~~~~~~~~~~~~~~~~~~~~~~~~~~~~( 3 + |\tan \theta_F|^2 ) \cosh \Delta \Gamma  ) \notag\\
&A_{o,2\omega} =  e^{ -\Gamma} |t_{xx}|^2  (  Im( \tan \theta_F)  \sin \Delta \beta  + \notag \\
&~~~~~~~~~~~~~~~~~~~~~~~~~~~~~ Re( \tan \theta_F) \sinh  \Delta \Gamma  ) \notag\\
&A_{i,2\omega} =   - e^{ -\Gamma } |t_{xx}|^2    \sinh  \Delta \Gamma \\
&A_{o,4\omega} =   e^{ -\Gamma} |t_{xx}|^2 \frac{1}{2} Re(\tan \theta_F)  ( \cos  \Delta \beta - \cosh \Delta \Gamma ) \notag \\
&A_{i,4\omega} = e^{ -\Gamma} |t_{xx}|^2 \frac{1}{4}  ( |\tan
\theta_F|^2 - 1 ) ( \cos \Delta \beta - \cosh \Delta \Gamma )
\notag
\end{align}.

$A_{i,2\omega}$, $A_{o,2\omega}$, $A_{i,4\omega}$, $A_{o,4\omega}$
are the second and fourth harmonic signals on the output detector
that are in-phase (i) and out-of-phase (o) with the original
harmonic signal that results when no sample is present or when the
magnetic field is zero. By taking various ratios of the measured
quantities in Eq.\,\ref{eq:detectorsignaHS}, one may extract the
quantities Re($\tan  \theta_F$) and Im($\tan \theta_F$).

In many circumstances, $t_{xy}$ is zero in zero magnetic field and
remains small even up to the maximum magnetic field such that
$|t_{xy}|^2 \ll |t_{xx}|^2$. In this small angle limit, the
Faraday angle may be obtained from Eq.\,\ref{eq:detectorsignaHS}:

\begin{align}\label{eq; RotWP-HS-smallangle}
&Re( \theta_F)_1  =   - \frac{1}{2} \frac{ A_{o, 4 \omega}}{ A_{i,4 \omega}} \notag \\
&Re( \theta_F)_2  =   \frac{1}{2} \frac{ \cos \Delta \beta + 3
\cosh  \Delta \Gamma}{\cos  \Delta \beta - \cosh \Delta \Gamma}
\frac{ A_{o,4 \omega}}{ R_{dc} } \notag \\
&Im( \theta_F)_1  =   \frac{- Re( \theta_F) \sinh  \Delta \Gamma}{ \sin  \Delta \beta} + \notag \\
&~~~~~~~~~~~~~~~~~~ \frac{1}{4} \frac{ \cos \Delta \beta + 3 \cosh \Delta \Gamma}{\sin \Delta \beta}  \frac{ A_{o,2 \omega}}{R_{dc} } \notag \\
&Im( \theta_F)_2  =   \frac{- Re( \theta_F) \sinh  \Delta \Gamma}{ \sin  \Delta \beta} +\notag \\
&~~~~~~~~~~~~~~~~~~ \frac{1}{4} \frac{ \cosh \Delta \Gamma - \cos \Delta \beta}{\sin \Delta \beta}  \frac{ A_{o,2 \omega}}{ A_{i,4 \omega} } \notag \\
\end{align}

The preferred ratios for evaluating the Faraday angle involve the
second and fourth harmonic amplitudes. The corrections involving
the detector frequency dependence become more sever for the case
of $R_{dc}$ due to the comparatively high chopping frequency.

\subsubsection{Polarizer or waveplate as a sample}

Inserting a polarizer at an angle $\alpha$ with respect to the
analyzer in place of the sample, and replacing the rotationally
invariant sample in Eq.\,\ref{eq:shorthanddetectorsignaHS} with
$\langle L|SP( \alpha )|L \rangle =
 \langle L|R( \alpha)|L \rangle \langle L|P|L \rangle \langle L| R( -  \alpha ) |L \rangle$
 yields the following detector signal:
\begin{align}\label{eq:polarizerAsSampleDetSigs}
&R_{dc}= \frac{e^{ -\Gamma}}{4} \cos^2 \! \alpha \; (\cos 2 \alpha \cos  \Delta \beta  + (2 + \cos 2 \alpha ) \cosh  \Delta \Gamma ) \notag\\
&A_{o,2\omega}=  -\frac{e^{ -\Gamma }}{2} \cos^2\! \alpha \; \sin 2 \alpha \sinh  \Delta \Gamma \notag \\
&A_{i,2\omega}= -  \frac{e^{ -\Gamma }}{2} \cos^2 \! \alpha \; (1+ \cos 2 \alpha) \sinh  \Delta \Gamma  \notag \\
&A_{o,4\omega}= - \frac{e^{ -\Gamma}}{4} \cos^2 \! \alpha \; ( \cos  \Delta \beta  - \cosh \Delta \Gamma ) \sin 2 \alpha \notag \\
&A_{i,4\omega} = - \frac{e^{ -\Gamma}}{4} \cos^2 \! \alpha \; (
\cos \Delta \beta  - \cosh \Delta \Gamma ) \cos 2 \alpha
\end{align}
Notice that $A_{o,4\omega}/A_{i,4\omega}=\tan 2\alpha$.

Inserting a waveplate in place of the sample defined by a
retardance and differential absorption given by $\Delta \beta
^\prime$ and $\Delta \Gamma ^\prime$, respectively, whose ordinary
axis is at an angle $\alpha$ with respect to the analyzer yields
the following transfer function
\begin{equation} \label{eqn; sample waveplate}
\langle L|SWP ( \alpha )|L \rangle =
 \langle L|R( \alpha)|L \rangle \langle L|WP|L \rangle
 \langle L| R( -\alpha ) |L \rangle \\
\end{equation}

Replacing $\langle L|S_0|L \rangle$ with $\langle L| SWP ( \alpha
)|L \rangle$ in Eq.\,\ref{eq:shorthanddetectorsignaHS} yields
detector amplitudes in terms of $\Delta \beta ^\prime$, $\Delta
\Gamma ^\prime$, $\alpha$, $\Delta \beta$, and $\Delta \Gamma$.

We make the following definition:
\begin{equation} \label{eqn; Fresnel sample waveplate}
\langle L|SWP ( \alpha )|L \rangle \equiv
\begin{pmatrix}
    t_{xx} ^\prime  &   t_{xy} ^\prime \\
    t_{yx} ^\prime  &   t_{yy} ^\prime
\end{pmatrix}
\end{equation}
With this definition, $t_{xy} ^\prime = t_{yx} ^\prime$, and if
$\Delta \Gamma^\prime$ is sufficiently small, then $t_{yy} ^\prime
\approx t_{xx} ^\prime$. Under these circumstances, the Faraday
angle may be derived from the sample waveplate parameters:
\begin{align}\label{eq:ReImFA-sampleWP}
Re( \theta_F) =&  \frac{1}{2}  (1 - \frac{|t_{xy}'|^2}{|t_{xx}'|^2})  \frac{ A_{o, 4 \omega} }{ A_{i,4 \omega t}} \notag\\
Im( \theta_F) =& - \frac{Re( \theta_F) \sinh \Delta \Gamma}{ \sin  \Delta \beta} + \notag\\
& \frac{1}{4}  (1-\frac{|t_{xy'}|^2}{|t_{xx}'|^2}) \frac{ \cos \Delta \beta- \cosh \Delta \Gamma }{\sin \Delta \beta} \frac{ A_{o,2 \omega}}{ A_{i,4 \omega}  } \notag\\
\end{align}
where $|t_{xy}'|^2/|t_{xx}'|^2$ is given by
\begin{equation}
 \frac{ \frac{1}{2}(\cosh \Delta \Gamma ^\prime -\cos \Delta \beta ^\prime )
\sin^2 2 \alpha}
 {e^{-\Delta \Gamma ^\prime} \cos^4  \alpha+e^{\Delta \Gamma ^\prime}
\sin^4 \alpha+\frac{1}{2} \cos \Delta \beta ^\prime \sin^2 2
\alpha}
\end{equation}

\subsection{Faraday angle from a Drude thin film sample}
A Drude response in magnetic field is derived by using the
classical force equation $q \vec{E}  + \frac{q}{c} \vec{v} \times
\vec{B} - \frac{m}{\tau} \vec{v} = m \frac{d \vec{v}}{d t}$
together with the definition of current $\vec{J} = n q \vec{v} =
\tilde{\sigma} \vec{E}$. The conductivity tensor is diagonal when
represented in the circular polarization basis:
\begin{equation}\label{eqn; sigma pm}
\langle C | \sigma | C \rangle =
\begin{pmatrix} \frac{n q^2}{m} \frac{1}{\gamma -  \mathit{i}
(\omega - \omega_c)}
& 0\\
0 & \frac{n q^2}{m} \frac{1}{\gamma -  \mathit{i} (\omega +
\omega_c ) }
\end{pmatrix}
 \equiv \begin{pmatrix}
 \sigma_+  & 0\\
 0         & \sigma_-
 \end{pmatrix}
\end{equation}
Since the matrix relating $\vec{J}$ and $\vec{E}$ is diagonal, the
complex Fresnel transmission coefficient for a plane wave at
normal incidence from vacuum through a thin conducting film
($\lambda \gg$ film thickness) on a semi-infinite substrate of
index n can be written as:
\begin{equation}\label{eqn; tpm}
t_{\pm}   =  \frac{ 2 } {1 + n + Z_0 \sigma_{\pm}d}
\end{equation}
where $Z_0 = 377 \; \Omega/\square$ is the impedance of free
space, and $d$ is the thickness of the film.

Utilizing the following conversions,
\begin{equation}\label{eqn; conductivity and t circ to lin}
\begin{split}
\sigma_{ \pm } &= \sigma_{xx} \pm \mathit{i} \; \sigma_{xy}
\\
t_{ \pm } &= t_{xx} \pm \mathit{i} \; t_{xy}
\end{split}
\end{equation}
and combining Eq.\,\ref{eqn; sigma pm} and \ref{eqn; tpm} gives:
\begin{equation}\label{eqn; Drude complex FA}
\begin{split}
\tan \theta_F & = - \frac{t_{xy}}{t_{xx}}\; = -\frac{\kappa \omega
_c}{(\gamma -i \omega ) (\gamma +\kappa -i \omega )+\omega _c^2}
\end{split}
\end{equation}
where $\kappa = \frac{n q^2}{m} Z^\prime$, $\omega_c = \frac{q \;
B}{m \; c}$, $Z^\prime = d \; Z_0 / (n + 1)$, $Z_0 = 377 \; \Omega
/ \square$, $n$ is the index of the substrate, and d is the film
thickness. $\kappa$ is a frequency which characterizes the
radiation damping of the 2-DEG, a plasma mode which takes into
account the reduction of the electric field inside the medium.

Two-dimensional electron gas heterostructures are usually
characterized in terms of the electron mobility and electron
density. The mobility can be defined as:
\begin{equation}\label{equ; mobility}
\mu \equiv \frac{q}{m} \frac{1}{\gamma}
\end{equation}
Inverting Eq.\,\ref{equ; mobility} and substituting $\gamma$ into
Eq.\,\ref{eqn; Drude complex FA} gives the Faraday angle in terms
of mobility.

\subsection{Converting from Faraday angle to Hall angle in the thin film limit}

The Hall angle is defined as:
\begin{equation}\label{eq; Hall angle def}
\tan \theta_H \equiv \frac{\sigma_{xy}}{\sigma_{xx}}
\end{equation}
Assuming a rotationally invariant sample in the thin-film limit,
and converting Eq.\,\ref{eqn; defn Faraday Rotation} to the
circular basis via Eq.\,\ref{eqn; conductivity and t circ to lin}
and using Eq.\,\ref{eqn; tpm} gives:
\begin{equation*}
\tan \theta_F  = \mathit{i} \; \frac{t_+ - t_-}{t_+ + t_-} =
\mathit{i} \; Z^\prime \; \frac{\sigma_- - \sigma_+}{2 + Z^\prime
(\sigma_+ + \sigma_-)}
\end{equation*}
Converting the conductivity back into the linear polarization
basis (Eq.\,\ref{eqn; conductivity and t circ to lin}) and
simplifying gives:
\begin{equation*}
\tan \theta_F = Z^\prime \; \frac{\sigma_{xy}}{1 + Z^\prime
\sigma_{xx}} =  \frac {1} {(Z^\prime \sigma_{xx} d)^{-1} + 1} \;
\frac{\sigma_{xy}}{\sigma_{xx}}
\end{equation*}

Substituting the definition of the Hall angle (Eq.\,\ref{eq; Hall
angle def}) and rearranging yields the desired result:
\begin{equation}\label{eq; Faraday To Hall Angle Conversion}
\tan \theta_H = \Big (1+ \frac{n + 1}{Z_0 \; \sigma_{xx} \; d}
\Big ) \tan \theta_F
\end{equation}

%

\end{document}